\documentstyle[aps,prl]{revtex}
\RequirePackage{graphicx}
\begin{document}

{\normalsize
\title{Peierls transition with acoustic phonons and
twist deformation in carbon nanotubes}
\author{Marc~Thilo~Figge, Maxim~Mostovoy, and
Jasper~Knoester
}
\address{Institute for Theoretical Physics and
Materials Science Center\\
University of Groningen, Nijenborgh 4, 9747 AG Groningen,
The Netherlands}
\date{\today}
\twocolumn
\maketitle
%
%
%

\begin{abstract}
We consider the Peierls instability due to the interaction of
electrons with both acoustic and optical phonons.  We suggest
that such a transition takes place in carbon nanotubes with small
radius.  The topological excitations and the temperature
dependence of the conductivity resulting from the
electron-lattice interactions are considered.
\end{abstract}


\medskip

\noindent PACS numbers: 63.20.Kr, 63.70.+h, 05.45.Yv, 72.80.Rj

\medskip


It is well-known that half-filled conducting chains are unstable
against a doubling of the unit cell, where the bond length
between neighboring lattice sites alternates along the chain
(dimerization).  The resulting alternation of the electron
hopping amplitudes opens a gap in the electron spectrum, turning
the system into a semiconductor (Peierls
instability)~\cite{Peierls3055}.  The phonons relevant for this
instability have wave vector $q \sim \pi/a$ ($a$ is the distance
between neigboring lattice sites).  These phonons backscatter
electrons from the left part of the one-dimensional Fermi surface
(wave vector $k \sim -\frac{\pi}{2a}$) to the right one ($k \sim
\frac{\pi}{2a}$) and vice versa.  It is usually assumed that
above the phase transition temperature $T_c$ these phonons have a
finite frequency and we will, therefore, refer to them as optical
phonons.  Due to the mixing with the low-energy electron-hole
excitations, these phonons soften and at $T_c$ their frequency
vanishes (giant Kohn anomaly)~\cite{Rice73}.

In this Letter we show that acoustic phonons (i.e.  phonons of
small wave vector $q$ and small frequency $\omega_a(q) = v_0
|q|$) may lead to a similar instability and that in the presence
of both optical and acoustic phonons, the acoustic ones are the
first to soften, independent of the coupling strengths involved.
The acoustic component immobilizes the topological excitations
(kinks) over the ordered state, while the softening leads to a
sharp increase of the electrical resistivity just above $T_c$.

At first sight, acoustic phonons cannot open an electronic gap,
as they only have small wave vectors, which does not allow for
backscattering.  This argument no longer holds, however, in
lattices where the unit cell already contains two sites above
$T_c$ (where the electron hopping amplitude is still uniform
along the chain), as in that situation the phonon wave vector is
only conserved up to a multiple of $\frac{\pi}{a}$.  For
instance, due to its zigzag structure the undistorted {\em
trans}-polyacetylene chain has lattice period $2a$ (Fig.~1(a)).
In Fig.~1(b) it is shown how a small-$q$ distortion perpendicular
to the chain leads to bond-length alternation, which results in
backscattering of electrons.  We note that in the strictly
one-dimensional Su-Schrieffer-Heeger model~\cite{Heeger88} of
{\em trans}-polyacetylene, this acoustic backscattering cannot
occur.

Another example is provided by an armchair single-wall carbon
nanotube (CNT) (Fig.~1(c)).  The undistorted armchair CNT has two
half-filled one-dimensional electronic bands and is
metallic~\cite{Saito98}.  Its unit cell contains two inequivalent
carbon atoms, which gives rise to backscattering of electrons off
acoustic phonons.  The relevant acoustic phonons are the
so-called twistons.  It is easy to see that a uniform twist of
the armchair CNT results in an alternation of the carbon-carbon
bond lengths along the nanontube (Fig.~1(d)), which opens a gap
in the spectrum of electron excitations.  The scattering of
electrons on twistons has recently been used to explain the
linear temperature dependence of the resisitivity of a
single-wall CNT~\cite{Kane98}.

%
%
\begin{figure}[]
\includegraphics[width=5.5cm]{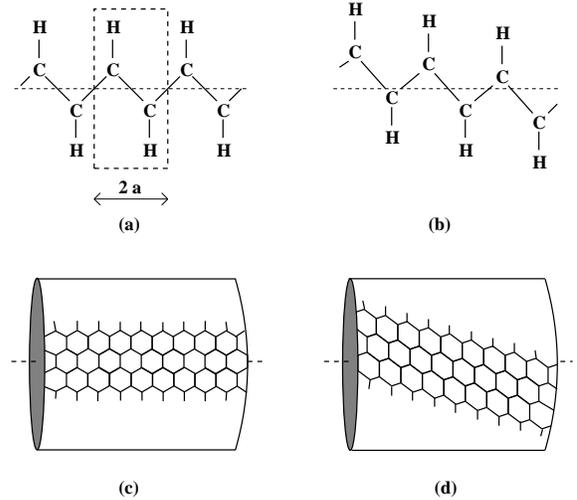}

\bigskip

\caption[]{\label{fig1} The zigzag structure of {\it trans}-polyacetylene
with two carbon atoms per unit cell, without lattice distortion (a) and 
with a small-wave-vector (acoustic) deformation perpendicular to the
chain (b).  Similarly, the undistorted armchair CNT (c) consists
of connected zigzag chains in the tube direction.  A twist
deformation of the CNT leads to an out of phase alternation of
the bond lengths in neighbouring chains.}
\end{figure}

The Hamiltonian of our model has the form,
\begin{eqnarray}
H &=& \sum_{n\sigma}
\int\!\!dx
\Psi^{\dagger}_{n\sigma}\Bigg\{
\frac{v_F}{i} \hat{\sigma}_3 \frac{\partial}{\partial x}
+ (\Delta_o + \Delta_a) \hat{\sigma}_1\Bigg\}\Psi_{n\sigma}
\nonumber\\
&+& H_{lat}[\Delta_o] + H_{lat}[\Delta_a],
\label{H}
\end{eqnarray}
where the first term describes the kinetic energy of electrons
close to the Fermi points and the backscattering of the electrons
due to the lattice modulation induced by both optical and
acoustic phonons.  Here, $\hat{\sigma}_1$ and $\hat{\sigma}_3$
are the Pauli matrices and we set $\hbar=1$.  The spinor
$\Psi_{n\sigma}={\psi_{Rn\sigma} \choose \psi_{Ln\sigma}}$
describes electrons moving to the right/left with the Fermi
velocity $v_F$.  We have omitted scattering contributions
that leave the electron on the same side of the Fermi surface ($R
\rightarrow R$ and $L \rightarrow L$), as these processes do not
lead to an instability.  The index $n=1,\ldots,N_b$ denotes the
electron band.  The second and the third terms in Eq.(\ref{H})
describe the energy of the optical $(i=o)$ and acoustic $(i=a)$
phonons,
\begin{equation}
H_{lat}[\Delta_{i}] =
\frac{N_b}{\pi \lambda_{i} v_F}\int\!\! dx\,\Delta_{i}^2 +
\frac{1}{2\rho_{i}}\int\!\! dx\,\pi_{i}^2,
\label{Hlat}
\end{equation}
where the dimensionless coupling constant $\lambda_{i}$ absorbs a
factor $N_b$ for the number of electron bands, and $\rho_{i}$ is
the mass density.
The momentum density $\pi_{i}$ is related to the phonon amplitude $u_i$ by:
$\pi_{i}(x,t) =\rho_{i} \partial u_{i}(x,t)/\partial t$.

In the continuum description of the optical and acoustic phonons
coupled to electrons, $\Delta_i$ is related to the phonon amplitude
$u_i$ by
\begin{equation}
\left\{
\begin{array}{ll}
\Delta_{o} \propto u_o(x,t) & \mbox{for optical phonons}, \\ \\
\Delta_{a} \propto \frac{\partial}{\partial x} u_a(x,t) &
\mbox{for acoustic phonons}.
\end{array}
\right.
\label{Deltau}
\end{equation}
The difference between the optical and acoustic phonons is due to
the fact that the ``order parameter'' $\Delta$ incorporates also
the electron-phonon coupling constant, as is clear from
Eq.~(\ref{H}).  While the coupling $g_{a}(q)$ to the acoustic
phonon with wave vector $q$ is proportional to $q$, the coupling
to the optical phonon, $g_{o}(q)$, is approximately constant.  At
the same time, Eq.~(\ref{Deltau}) gives the linear dispersion for
acoustic phonons, $\omega_{a}=v_0 |q|$, and a finite frequency
$\omega_{o}$ for optical phonons.  We note that, though the
coupling to acoustic phonons $g_{a}(q)$ is small for small $q$,
the actual strength of the interaction is given by the
dimensionless coupling $ \lambda_{i} \propto g_{i}^2/(v_F
\omega_{i}^2)$, which is finite both for optical and acoustic
phonons.

For $\Delta_a = 0$ and $N_b=1$ the Hamiltonian Eq.~(\ref{H})
coincides with the Hamiltonian of the TLM
model~\cite{Takayama80}, which is the continuum version of the
Su-Schrieffer-Heeger model~\cite{Heeger88} of {\em
trans}-polyacetylene with one electron band.  On the other hand,
for $\Delta_o = 0$ and $N_b = 2$, the Hamiltonian Eq.~(\ref{H})
is equivalent to the Hamiltonian describing the electron-twiston
interactions in the armchair CNTs~\cite{Kane97,Kane98}.


{\it Phase transition.}---First we consider the temperature
dependence of the optical and acoustic phonon frequencies for $T
> T_c$.  The bare phonon propagation $\hat{D}_0$ and the the
coupling $\hat{V}$ to electrons is formally written in terms of
$2\times 2$-matrices,
\[
\hat{D}_{0}=\left(
\begin{array}{cc}
D_a^0 & 0\\
0 & D_o^0
\end{array}
\right)\;\;\;{\rm and}\;\;\;
\hat{V}=\left(
\begin{array}{cc}
g^{}_a\, g_a^{\ast} & g^{}_a\, g_o^{\ast}\\
g^{}_o\, g_a^{\ast}& g^{}_o\, g_o^{\ast}
\end{array}
\right)P\;,
\]
where $\hat{D}_0$ contains the bare acoustic and optical phonon
propagators $D^{0}_a$ and $D^{0}_o$.  The coupling $\hat{V}$ of
the phonons to electrons has two effects: It mixes the bare
optical and acoustic phonon modes and renormalizes their
frequencies due to an electron-hole excitation that is described
by the bare vacuum polarization $P=P(q, \omega,
T)$~\cite{Mahan81}.  Within the random phase approximation, the
propagation of dressed phonons is described by a matrix $\hat{D}$
that solves the Dyson equation $\hat{D} = \hat{D}_0 + \hat{D}_0
\hat{V} \hat{D}$.  The renormalized phonon frequencies ${\tilde
\omega}_o(q)$ and ${\tilde \omega}_a(q)$ are found from the poles
of ${\rm det}(\hat{D})$, or equivalently by solving ${\rm
det}({\hat{D}_0}^{-1}-\hat{V})=0$.  Using that ${\tilde
\omega}_a(q)\ll T$ and ${\tilde
\omega}_a(q)\ll\tilde{\omega}_o(q)$, we eventually find:
\begin{equation}
{\tilde \omega}^2_a(q) =
\omega^2_a(q) \,\frac{1-(\lambda_o+\lambda_a)\ln(\gamma W/\pi T)}
{1-\lambda_o\ln(\gamma W/\pi T)}
\label{newacu}
\end{equation}
for the acoustic phonon and, assuming that ${\tilde \omega}_o(q)\ll T$,
\begin{equation}
{\tilde \omega}^2_o(q) =
\omega^2_o(q)\left(1-\lambda_o\ln(\gamma W/\pi T)\right)
\label{newopt}
\end{equation}
for the optical phonon.  Here, $W$ is the energy cut-off of the
order of the electron band width and $\gamma= 1.781072...$ is
Euler's constant.

We thus see, that the expression for the renormalized optical
frequency is independent of $\lambda_a$ and is the same as in the
absence of the coupling to acoustic phonons.  On the other hand,
the renormalized acoustic phonon frequency depends on the sum of
$\lambda_a$ and $\lambda_o$.  As a result, the acoustic phonons
``soften'' first, at the critical temperature given by
\begin{equation}
T_c\;=\;\frac{\gamma}{\pi}
W\exp\left(-\frac{1}{\lambda_a+\lambda_o}\right).
\label{transc}
\end{equation}
As ${\tilde \omega}_a(q = 0) = 0$ at all temperatures, the
``softening'' in this case means vanishing of the acoustic phonon
velocity at $T=T_c$.  Thus, no matter how much the optical
coupling constant is larger than the acoustic
coupling constant, it is always the velocity of the
acoustic phonon that becomes zero at the transition temperature,
whereas the optical phonon frequency stays finite
at $T = T_c$.  This is a consequence of the mixing of the optical
and acoustic phonons due to their interactions with electrons,
which results in a repulsion between the frequencies of the two
modes.  As a result, the optical and acoustic branches can
never cross and the singularity at $T_c$ always occurs in the
lower, i.e. acoustic, branch.  A similar effect takes place in
some ferroelectrics, in which the sound velocity vanishes at the
transition temperature because of the mixing of the soft
mode, describing the ferroelectric displacement of ions, to
acoustic phonons \cite{Brody68}.  We also note that, at first
sight, Eq.~(\ref{transc}) resembles the result for the Peierls
temperature obtained in Ref.\CITE{Huang96}.  It should be kept in
mind, however, that in that paper the various coupling constants
$\lambda_i$ correspond to contributions from scattering within
different electron bands and are not associated with the presence
of several phonon modes.  The additive effect of the number of
electron bands is implicit in our result through the fact that
both $\lambda_o$ and $\lambda_a$ are proportional to $N_b$.

Next we discuss the ordered state below $T_c$.  Because the
lattice distortions, corresponding to the optical and acoustic
phonons, are coupled due to electron-phonon interactions, both
order parameters, $\Delta_a(x)$ and $\Delta_o(x)$, should appear
below the transition temperature.  In the mean field treatment of the
lattice one has to minimize the total free energy of the model
Eq.(\ref{H}) with respect to the two order parameters.  The
solutions of the resulting self-consistency equations have the
following properties: (i) The sum of the optical and acoustic
order parameters, $\Delta(x)=\Delta_o(x)+\Delta_a(x)$, satisfies
the same Bogoliubov-de Gennes equations as the order parameter of
the TLM model \cite{Takayama80} with a single phonon mode,
but with the coupling constant $\lambda = \lambda_o+\lambda_a$; (ii)
The optical $(i=o)$ and acoustic $(i=a)$ order parameters are proportional
to $\Delta(x)$:
\[
\Delta_{i}(x)=\frac{\lambda_{i}}{\lambda} \Delta(x).
\]

Thus, the solutions of the TLM model can be used to study the
ordered state in our model.  In particular, the order parameter
for the homogeneous solution $\Delta(x) = \Delta_0$ at zero
temperature has the value $\Delta_0 = \frac{\pi}{\gamma} T_c$,
with $T_c$ as in Eq.~(\ref{transc}).

The homogeneous optical order parameter, $\Delta_o$, is a
``frozen'' phonon mode, corresponding to a uniform dimerization
of the chain.  On the other hand, for acoustic phonons, the
constant $\Delta_a$ corresponds to the linearly growing amplitude
of ionic displacements: $u_a(x) \propto \Delta_a x$.  Such a
lattice distortion is not a ``frozen'' phonon mode, as it
corresponds to large deviations of ions from their equilibrium
positions in the high-temperature phase.  In the case of CNTs,
$\Delta_a={\rm const.}$ corresponds to a uniform twist of the CNT
(see Fig.~\ref{fig1}).  Such a twist was, in fact, recently
observed using STM~\cite{Clauss98}.


{\it Topological Excitations.}---In the SSH model, where
electrons interact with one optical phonon mode, domain walls
(kinks) in the order parameter, corresponding to a change of sign
of the lattice dimerization along the chain, constitute an
interesting class of excitations~\cite{Heeger88}.  In the
continuum model, the analytical expression for the kink is given
by~\cite{Takayama80}:
\begin{equation}
\Delta(x) = \Delta_0\tanh \frac{x}{\xi_0},
\label{kinksol}
\end{equation}
where $\xi_0=\frac{v_F}{\Delta_0}$ is the correlation length.

The kink Eq.~(\ref{kinksol}) also is a consistent solution of the
Bogoliubov-de Gennes equations for the total order parameter in
our model with both optical and acoustic phonons.  For this kink,
the optical and acoustic lattice distortions take the form:
\begin{equation}
\left\{
\begin{array}{lcr}
u_o(x) & = & \bar{u}_o \tanh\frac{x}{\xi_0},\\ \\
u_a(x) & = &
\bar{u}_a\,\ln\cosh \frac{x}{\xi_0}.
\end{array}
\right.
\label{topex}
\end{equation}
Near the kink both the lattice dimerzation, described by
$u_o(x)$, and the derivative of $u_a(x)$ change sign (see
Fig.~\ref{fig2}).  In the case of a CNT, the latter corresponds
to the change of the sign of the twist-angle from
$-(\bar{u}_a/\xi_0)$ to $+(\bar{u}_a/\xi_0)$.

\bigskip

%
%
\begin{figure}[]
\includegraphics[width=8cm]{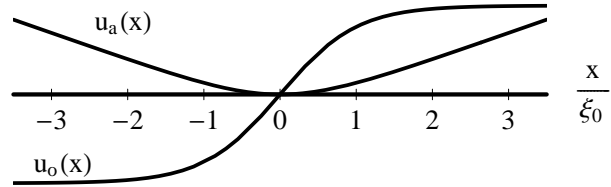}

\bigskip

\caption[]{\label{fig2} Schematic plot of the optical $(u_o(x))$ and
acoustic $(u_a(x))$ lattice distortions
Eq.~(\ref{topex}) corresponding to the topological excitation
Eq.~(\ref{kinksol}).}
\end{figure}

The creation energy of the kink in the model Eq.(\ref{H}) is the
same as for the kink in the TLM model~\cite{Takayama80}: $E_k =
N_b (2\Delta_0/\pi)$.  The dynamical properties of the kinks in
these two models are, however, quite different.  In the TLM model
the kink can propagate with velocity $v$ along the chain, without
changing its profile.  This results from the independence of the
kink energy of its position and the fact that the kinetic energy
density of the moving kink, $u_o(x,t) = \bar{u}_o
\tanh\left(\frac{x - vt}{\xi_0}\right)$ decays exponentially at
distances larger than $\xi_0$ from the kink.  Thus, the mass of
the kink (defined by equating the kinetic energy of the moving
kink to $\frac{1}{2} M v^2$) is finite and for {\it
trans}-polyacetylene it was estimated to be $\sim 6
m_e$~\cite{Heeger88}.  On the other hand, the motion of the kink
in the model with acoustic phonons would result in a constant
kinetic energy at distances larger than $\xi_0$ from the kink, as
follows from the substitution of $u_a(x)$ in Eq.(\ref{topex}) by
$u_a(x - vt)$.  Thus, the mass of such a kink is proportional to
the chain length $L$.  This relates to the fact that a
translation of the kink configuration Eq.(\ref{topex}) changes
the coordinates of the chain ends, so that a shift of the kink
induces a motion of the entire chain.  Thus, in our model,
isolated kinks cannot propagate along the chain.  Kinks can
propagate without affecting the chain ends only together with
antikinks, the corresponding time-dependent lattice configuration
being $u_R(x, t) = u_a(x-vt) - u_a(x+R-vt)$ [cf.
Eq.~(\ref{topex})].  The mass of such a pair is proportional to
the distance $R$ between the kink and antikink.


{\it Electrical conductivity.}---Next we consider the temperature
dependence of the electrical conductivity, $\sigma(T)$, above
$T_c$.  For $T \ll\tilde{\omega}_o$, the contribution of the
optical phonons to the conductivity can be neglected.  For
$\lambda_a \ll \lambda_o$ this inequality may not be fulfilled
close to $T_c$, as the renormalized optical phonon frequency at
$T_c$, ${\tilde\omega}_o = \omega_o
\sqrt{\lambda_a/(\lambda_a+\lambda_o)}$, may then be small.
However, as we argue below, for CNTs, $\lambda_a \sim \lambda_o$,
in which case there is no dramatic softening of the optical
phonon and the conductivity close to $T_c$ is also dominated by
the electron backscattering off acoustic phonons.

The electrical conductivity is given by $\sigma(T)=(-4 e^2
v_F/\pi)\int dk\,\tau_k (\partial n_F(v_F k)/\partial k)$, where
$n_F$ is the Fermi distribution and $\tau_k$ is the transport
lifetime, which depends on the electron wave vector $k$.  
Using Fermi's Golden Rule and accounting 
for the renormalization Eq.(\ref{newacu}) of the acoustic phonon 
frequency, we obtain
\begin{equation}
\tau_k = \frac{2}{\pi \lambda_a T}
\left(\frac{\tilde{\omega}_a(2k)}{\omega_a(2k)}\right)^2\;,
\label{trlifetime}
\end{equation}
where we have put $k_B=1$.
%
%
\begin{figure}[]
\includegraphics[width=8cm]{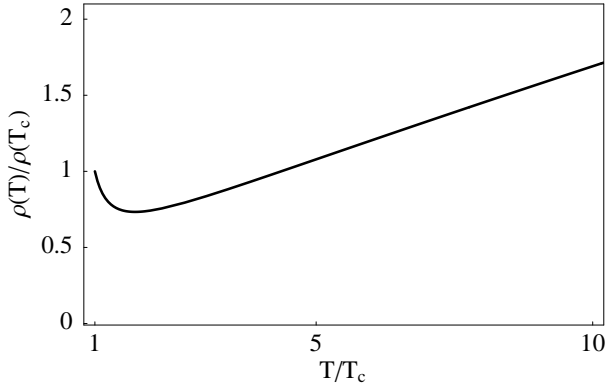}

\smallskip

\caption[]{\label{fig3} The electrical resistivity
$\rho=1/\sigma$ as a function of temperature.  The
electron-phonon coupling strengths are chosen to be $\lambda_a =
\lambda_o = 0.05$.}
\end{figure}

Fig.~\ref{fig3} shows the temperature dependence of the
electrical resistivity, $\rho(T)=1/\sigma(T)$, calculated for $T
\ge T_c$ using Eq.(\ref{trlifetime}).  At temperatures $T\gg
T_c$, the resistivity decreases linearly with temperature,
$\rho(T) \propto \lambda_a T$, as was also found in
Refs.~\cite{Kane98,Jishi93}.  However, close to $T_c$, due to the
vanishing of the acoustic phonon velocity at the critical
temperature, the resistivity strongly increases up to some finite
value, $\rho(T_c)$.  This behavior is very similar to the one
observed for bundles of single-wall CNTs~\cite{Kane98}, where the
cross-over between the linear decrease and the sharp upturn of
the resistivity occurs at $T^{\ast}\sim 10-100$ K.

Usually, the Peierls instability is thought to be irrelevant
for CNTs, because the electron-phonon coupling constant is 
inversely proportional to the number of zigzag chains, $2N$,
in the armchair CNT, making $T_c$ negligibly small for 
$N\geq4$~\cite{Saito98}.
We note, however, that previously only the interaction with optical
phonons, {\em e.g.}, the deformations resulting in a Kekul\'{e}
or quinoid type structure of the CNT lattice, was
considered~\cite{Saito98}.  When we take into account the
scattering of electrons on acoustic phonons (twistons), we obtain
a higher value of the coupling constant $\lambda$ ($=\lambda_o +
\lambda_a$), which gives a much higher value of $T_c$.  In
particular, using typical parameters for graphite~\cite{Kane98},
we obtain $\lambda_a\simeq 0.23/N$.  Furthermore, if one assumes
that the electron hopping amplitudes and the spring constants
only depend on the carbon-carbon bond length, one finds
$\lambda_o \sim \lambda_a$.  For CNTs with $N=4$, which were
recently discussed in the context of possible technological
applications~\cite{Farajian99}, the critical temperature given by
Eq.~(\ref{transc}) is then $\sim 10$ K, instead of only a few mK
for the Peierls transition temperature without acoustic phonons.
We note, however, that the interaction of the CNTs with their
environement (other CNTs and the substrate) may, in principle,
suppress the Peierls instability.

The values of the electron-phonon coupling constants for a CNT
could, in principle, be obtained from the form of the resistivity
curve ($\lambda_a$ determines the slope of $\rho(T)$ at high
temperatures, whereas $T_c$ depends on $\lambda_a + \lambda_o$).
However, so far, such measurements have only been performed on
bundles of nanotubes, in which one cannot control the radius and
chirality of the CNTs.  Therefore, measuring the resistivity of a
single CNT would be very desirable.

{\it Conclusions.}---In this Letter we studied the Peierls
transition resulting from electrons interacting with both optical
and acoustic phonons.  We found that, independent of the
electron-phonon coupling constants, the acoustic phonon velocity
vanishes at the critical temperature, whereas the optical phonon
frequency remains finite.  This is different from the
conventional Peierls scenario, in which the optical phonon
softens at $T_c$.  We suggested that such a transition may take
place in CNTs of small radius, leading to a static twist below
$T_c$.  The temperature dependence of the resistivity calculated
within our model qualitatively agrees with experimental data on
CNTs.  
Unlike the topological excitations in {\it trans}-polyacetylene,
the kinks in our model are immobile even in isolated
chains and can propagate only in pairs.
Finally we note that, though we used CNTs as a specific example,
we believe that our results have a more general significance.

We gratefully acknowledge financial support by the
Stichting Fundamenteel Onderzoek der Materie (FOM).

\end{document}